\documentclass{article}

\usepackage{amssymb}

\usepackage{epsfig}

\usepackage{lscape}

\usepackage{graphicx,psfrag,amsmath}

\begin{document}

\def\beq#1\eeq{\begin{equation}#1\end{equation}}
\def\beql#1#2\eeql{\begin{equation}\label{#1}#2\end{equation}}

\def\bea#1\eea{\begin{eqnarray}#1\end{eqnarray}}
\def\beal#1#2\eeal{\begin{eqnarray}\label{#1}#2\end{eqnarray}}

\newcommand{\Z}{{\mathbb Z}}
\newcommand{\N}{{\mathbb N}}
\newcommand{\C}{{\mathbb C}}
\newcommand{\Cs}{{\mathbb C}^{*}}
\newcommand{\R}{{\mathbb R}}
\newcommand{\intT}{\int_{[-\pi,\pi]^2}dt_1dt_2}
\newcommand{\cC}{{\mathcal C}}
\newcommand{\cI}{{\mathcal I}}
\newcommand{\cN}{{\mathcal N}}
\newcommand{\cE}{{\mathcal E}}
\newcommand{\cA}{{\mathcal A}}
\newcommand{\xdT}{\dot{{\bf x}}^T}
\newcommand{\bDe}{{\bf \Delta}}

\def\ket#1{\left| #1\right\rangle }
\def\bra#1{\left\langle #1\right| }
\def\braket#1#2{\left\langle #1\vphantom{#2}
  \right. \kern-2.5pt\left| #2\vphantom{#1}\right\rangle }
\newcommand{\gme}[3]{\bra{#1}#3\ket{#2}}
\newcommand{\ome}[2]{\gme{#1}{#2}{\mathcal{O}}}
\newcommand{\spr}[2]{\braket{#1}{#2}}
\newcommand{\eq}[1]{Eq.\,(\ref{#1})}
\newcommand{\xp}[1]{e^{#1}}

\def\limfunc#1{\mathop{\rm #1}}
\def\Tr{\limfunc{Tr}}

\def\dr{detector }
\def\drn{detector}
\def\dtn{detection }
\def\dtnn{detection}

\def\pho{photon }
\def\phon{photon}
\def\phos{photons }
\def\phosn{photons}
\def\mmt{measurement }
\def\an{amplitude}
\def\a{amplitude }
\def\co{coherence }
\def\con{coherence}

\def\st{state }
\def\stn{state}
\def\sts{states }
\def\stsn{states}

\def\cow{"collapse of the wavefunction"}
\def\de{decoherence }
\def\den{decoherence}
\def\dm{density matrix }
\def\dmn{density matrix}

\newcommand{\mop}{\cal O }
\newcommand{\dt}{{d\over dt}}
\def\qm{quantum mechanics }
\def\qms{quantum mechanics }
\def\qml{quantum mechanical }

\def\qmn{quantum mechanics}
\def\mmtn{measurement}
\def\pow{preparation of the wavefunction }

\def\me{ L.~Stodolsky }
\def\T{temperature }
\def\Tn{temperature}
\def\t{time }
\def\tn{time}
\def\wfs{wavefunctions }
\def\wf{wavefunction }
\def\wfn{wavefunction} 
\def\wfsn{wavefunctions}
\def\wvp{wavepacket }
\def\pa{probability amplitude } 
\def\sy{system } 
\def\sys{systems }
\def\syn{system} 
\def\sysn{systems} 
\def\ha{hamiltonian }
\def\han{hamiltonian}
\def\rh{$\rho$ }
\def\rhn{$\rho$}
\def\op{$\cal O$ }
\def\opn{$\cal O$}
\def\yy{energy }
\def\yyn{energy}
\def\yys{energies }
\def\yysn{energies}
\def\pz{$\bf P$ }
\def\pzn{$\bf P$}
\def\pl{particle }
\def\pls{particles }
\def\pln{particle}
\def\plsn{particles}

\def\plz{polarization  }
\def\plzs{polarizations }
\def\plzn{polarization}
\def\plzsn{polarizations}

\def\sctg{scattering }
\def\sctgn{scattering}

\def\prob{probability }
\def\probn{probability}

\def\om{\omega} 

\def\hf{\tfrac{1}{2}}

\def\zz{neutrino }
\def\zzn{neutrino}
\def\zzs{neutrinos }
\def\zzsn{neutrinos}

\def\zn{neutron }
\def\znn{neutron}
\def\zns{neutrons }
\def\znsn{neutrons}

\def\csss{cross section }
\def\csssn{cross section}

\def\vhe{very high energy }
\def\vhen{very high energy}

\def\mult{multiplicity }
\def\multn{multiplicity}

\title{Cross Section to Multiplicity Ratios at
 Very High Energy}

\author{ M. M. Block\\
 Department of Physics and Astronomy \\
   Northwestern University,\    Evanston, Illinois 60208, USA \\
\\
L. Stodolsky, \\
Max-Planck-Institut f\"ur Physik
(Werner-Heisenberg-Institut)\\
F\"ohringer Ring 6, 80805 M\"unchen, Germany}

\maketitle

\begin{abstract}
 Recent data from the LHC makes it possible to examine an old
speculation that at \vhe  the total multiplicity and the \csss in
elementary \pl
interactions vary in parallel with \yyn. Using fits incorporating
the new data,  it appears that the ratios of the total, elastic,
and inelastic
\csssn s to the average \mult $N$ can in fact approach   constants
at very high \yyn. The approach to the limit is however quite slow
for the total and inelastic \csssn s 
and is not yet  reached at LHC \yysn. The elastic ratio
$\sigma^{el}/N$ at 7 TeV, however, is not far from its asymptotic
value.
\end{abstract}

\section{Introduction}
  Some time ago, one of us suggested 
\cite{leo}  that, at \vhe in elementary \pl collisions, the \csss
and the multiplicity should vary in the same way with \yyn. This
idea arises in a picture, inspired by calculations in (massive
photon) QED
\cite{yennie}, where the incoming \pl has an expanding radius
induced by a series of N independent emissions of secondary \plsn.
This leads to a random walk in the transverse dimension R, so that
one has $R\sim \sqrt{N}$, or using $\sigma \sim R^2$.

\beql{one}
\sigma \propto N\,.
\eeql
 The square root of the  constant of the proportionality would
have the interpretation of the step length in the random walk.
While in ref\,\cite{yennie} $N$ is the number of (massive) photons
emitted,
in hadron reactions we take it as proportional to the multiplicity,
which in
the following we also call $N$ (see section\,\ref{disc}).

 Although in
contemporary language one would not speak of QED for hadronic
interactions, the situation with QCD is not so much different, and
the idea
is  simple and general, so one may wonder if the relation does not
indeed hold.

 The picture is motivated \cite{leo}  by the  fact that coupling a
\pl to a light boson field will inescapably lead to a narrowing of
its purely elastic \sctg peak. This however implies \sctg at high
impact paramter,  showing that the boson field delocalizes the
\pln. Assuming this delocalization takes place in a series of
independent emissions, there is a diffusion in the transverse
dimension,  leading to $R\sim \surd N$. This occurs in the model of
ref\,\cite{yennie} as the mass of the boson field is reduced, and
we speculated there is an analogous behavior in \vhe hadron
physics \cite{gribov}.

\section{LHC Information}
That \eq{one} may  be true is suggested by two
aspects of  recently available information from the
LHC.

Firstly, good fits \cite{bh} to both the total and inelastic pp
\csss are possible with asymptotic
 $ \ln ^2\sqrt{s}$ behavior, where $\sqrt{s}$ is  the total cms 
(center of mass \sy) \yyn.

Secondly,  data available for $dN/dy$, the charged \pl
multiplicity
density in the rapidity $y$ \cite{cms}, show a rise with \yyn,
and this rise resembles a $\ln\sqrt{s}$ behavior, (Fig 1, below).
Such \mult information at the LHC is only available for central
rapidities, $y\approx 0$, at present.
But, one may
 estimate  the total multiplicity N by multiplying the central
$dN/dy$ by the 
the total rapidity interval $Y$ expected for the secondaries
(mostly pions):
\beql{nest}
N\approx\frac{dN}{dy}\bigg\vert_{y=0} \times Y\,,
\eeql
 $Y$ also grows as $\ln\sqrt{s}\,$, therefore suggesting $N\sim
\ln^2\sqrt{s}$. With both the \csss and $N$ showing the same
$\ln ^2\sqrt{s}\,$,
aymptotic growth, \eq{one} would hold.

\subsection{N Behavior} \label{nbhv}
\begin{figure}[h]
\includegraphics[width=\linewidth]{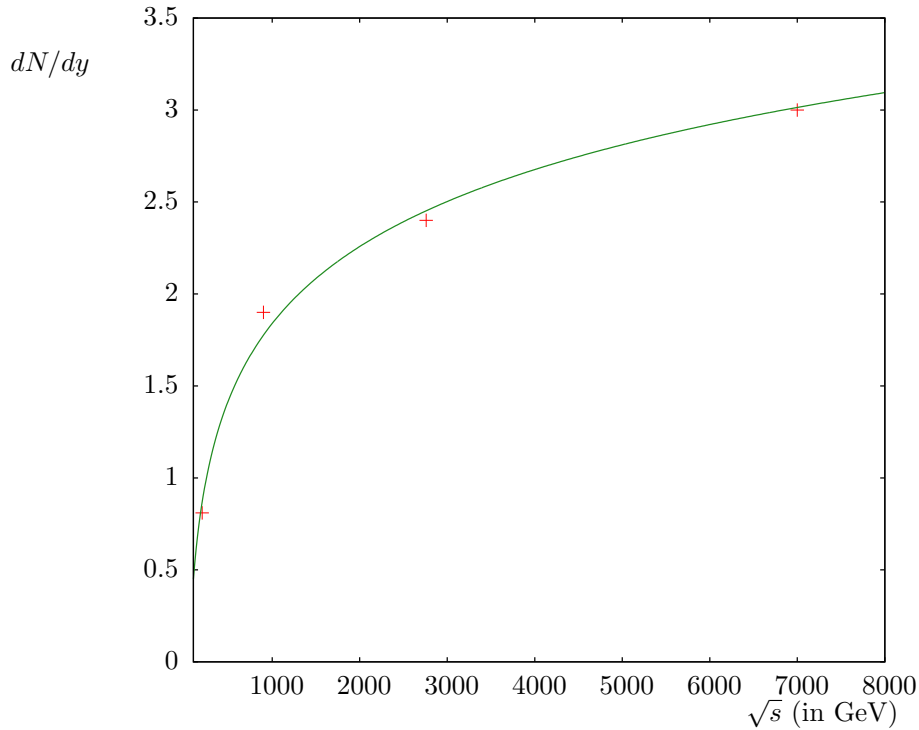}
\caption{ Fit to the \yy dependence, $\sqrt{s}$ in GeV, of the
central rapidity density
dN/dy for one charge of the pion, with  data points selected from
\cite{cms},
Fig 20. The fit
is of the form  $A \,\ln (\sqrt{s}/B)$, with A=0.60 and B=47 GeV}
\label{dndy}
\end{figure}

To substantiate these statements concerning the total \mult
we show in Fig\,\ref{dndy} a logarithmic fit to  central  $dN/dy$
data
selected from
 Fig 20 (top) of ref\,\cite{cms}. We use the data for charged
pions (one charge) .
There is considerable scatter in  the points at lower \yys and in
the
interest of dealing with a smooth curve, we simply use the 
point  at
 200 GeV, that appears  to connect smoothly to the
higher \yy CMS points.
A fit of form $A\, \ln \sqrt{s}/B$ yields A=0.60 and B=48 GeV. Thus
we will use
\beql{dd}
\frac{dN}{dy}\bigg\vert_{y=0}\approx 0.60\times (\ln (\sqrt{s}/GeV)
-3.9)
\eeql 
in our estimates. 
Clearly this selective choice of data is not useful for
estimating errors or uncertainties.
This result is for one charge of the pions, for ``all charged'' one
must mutiply
approximately by 2, and for total \mult including $\pi^0$'s by 3.
Kaons and
heavier \pls are at the 10\% level or less.

The various logarithmic expressions we have to  deal with are
of the form
$\ln  (\sqrt{s}/\mu)$ where $\mu$ is some mass scale. Since $\ln 
(\sqrt{s}/\mu)= \ln  (\sqrt{s}/GeV)- \ln (\mu/GeV)$, we can also
write
the expressions as $\ln  (\sqrt{s}/GeV) +constant)$, as in \eq{dd}.
This remark also
implies that in determining the asymptotic behavior  of the
 expressions, only the coefficient of the logarithmic terms
and not the scale $\mu$ enters.

\section{Asymptotic Ratios}

\subsection{Asymptotic \csssn s}\label{assig}  According to the
$c_2$ coefficients of Ref. \,\cite{bh} the asymptotic behavior of
the
pp inelastic \csss  is  $4 c_2^{inel} mb\times \ln
^2(\sqrt{s}/GeV)=
0.56\, mb\times \ln ^2(\sqrt{s}/GeV)$, while as expected for a
"black
disc"  one has $ 4 c_2\, mb \times \ln ^2(\sqrt{s}/GeV)=1.1\, mb
\times \ln ^2(\sqrt{s}/GeV)$ for the total \csssn. 

\subsection{Asymptotic \mult}\label{asmul}
 In the fit of Section \ref {nbhv} we arrived at $A=0.6$ as the
coefficient
of the logarithm for $dN/dy$ for a single charge of the pion.
 For $Y$  we anticipate  $Y=2\times
\ln (\sqrt{s}/\mu)$,  The  factor 2 is chosen so that with
$\mu=m_{proton}$, $Y$ is twice the proton rapidity in the cms. Thus
 multiplying by 3 for all pions, we will have 
\beql{allpi}
 N\approx 3.6 \times \ln ^2(\sqrt{s}/GeV) \,.
\eeql
for the  asymptotic  total \multn .

\subsection{Aysmptotic ratios}
Combining  sections \ref{assig} and  \ref{asmul} we have

\beql{asr}
\frac{\sigma^{in}}{N}\approx 0.16 {\rm \,mb}  ~~~~~~~~~~~~~~~~~ 
\frac{\sigma^{tot}}{N} \approx  0.31 {\rm \, mb} \,.
\eeql
for the asymptotic ratios. The black disk limit obtains, where
$\sigma^{elastic}/N=\sigma^{inel}/N$ and $\sigma^{tot}/N=2
\,\sigma^{in}/N$.

The units are naturally mb, or one might like to say, mb per pion.

\section{Approach to the Limit}
It is interesting to see how the limits \eq{asr} are
approached and how close the limits are at present \yysn. Since
even at
LHC \yys the black disk limit is remote and  $\sigma^{in}$ and
$\sigma^{tot}$ do not yet have the same
\yy dependence (see Fig 3 of ref\,\cite{bh}) the behavior of
$\sigma^{in}/N$ and $\sigma^{tot}/N$ and their difference
$\sigma^{el}/N$  will be somewhat different. 

\begin{figure}[h]
\includegraphics[width=\linewidth]{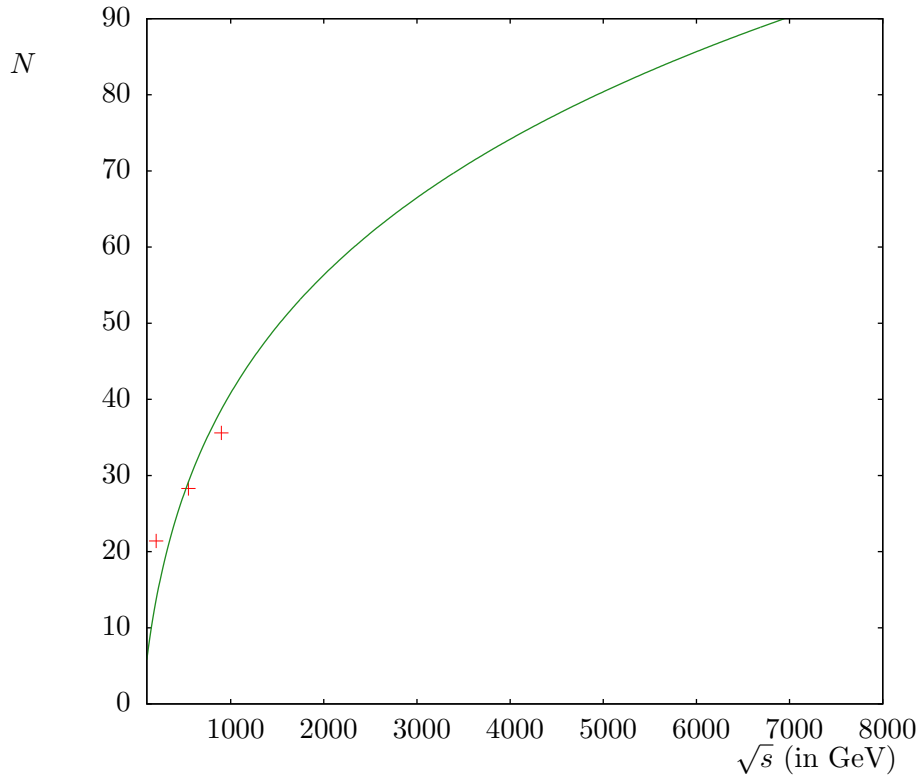}
\caption{ Plot of the estimate for the total charged \mult N, vs
$\sqrt{s}$ in
GeV, up to LHC \yysn. It is obtained  by
using \eq{nest} with $\frac{dN}{dy}\big\vert_{y=0}$ from experiment
(Fig\,\ref{dndy}) and $Y=2(\ln  (\sqrt{s/}GeV) +C)$, with $C=-1.3$,
as
obtained by fitting to UA5 data below 1 TeV (red crosses). For the
total
\multn, including $\pi^0$'s, these values should be muliplied by
3/2. }
\label{ntot}
\end{figure}

To examine the approach to the limit, we need not only the
coefficient of the $\ln ^2$ term but  
the actual total multiplicity at present LHC \yysn. The PDG tables 
\cite{pdg} give the total charged  \mult up to $\sqrt{s}$ of 
900\,GeV,
 using UA5 data. For the higher \yysn, we can make an estimate
using \eq{nest}.  For  $dN/dy$ we have the fit \eq{dd}. For $Y$ it
is now necessary to have the non-leading term, the
constant C  in $Y=2(\ln (\sqrt{s}/GeV)+ C)$. By requiring that
\eq{nest},
with \eq{dd} multiplied by 2, fits to the 
the UA5 points at 200, 546, and 900 GeV \cite{ua5} with N=21, 28
and 36 for the charged \mult we obtain $C \approx -1.3 $. 
Fig. \,\ref{ntot} shows  this fit with the UA5 points.

The negative value for $C$ indicates that the effective $Y$ is
somewhat
less than that between the incoming protons; this may be a
reflection of the
fact that $dN/dy$ falls off at large 
$y$  so that using its central value throughout, as we do,  tends
to give
an overestimate that must be compensated by a smaller $Y$
\cite{totema}.

 We shall thus use for the total pion \mult above 1000\,GeV 
\beal{nfit}
N&\approx 3 \times \bigl ((0.60\times (\ln (\sqrt{s}/GeV)
-3.9)\bigr)
\times 
\bigl(2(\ln (\sqrt{s}/GeV) -1.3) \bigr)\\
\nonumber
&=3.6\times \ln ^2(\sqrt{s}/GeV) -19\times \ln (\sqrt{s}/GeV) +18
\eeal

\begin{figure}[h]
\includegraphics[width=\linewidth]{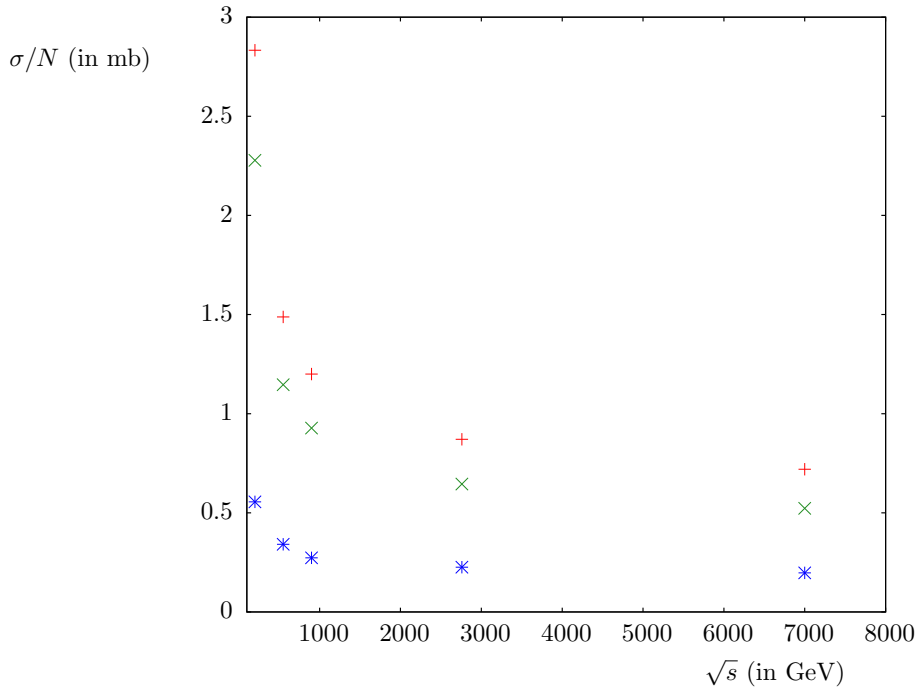}
\caption{ Ratios of the total (red +), inelastic (green x) and
 elastic (blue *) \csssn s to 
the total
pion \multn, in mb vs. $\sqrt{s}$ in GeV,  at LHC \yysn, using
the fits described in the text.
 According to \eq{asr} these ratios
should {\it asymptotically} approach $0.31$ mb,  $0.16$ mb, and
$0.16$ \,mb,
respectively.}
\label{ratios}
\end{figure}

 With this estimate for N and the fits for the $\sigma$
from ref\,\cite{bh} we obtain the  $\sigma/N$  values 
shown in Fig. \,\ref{ratios}.

It is appears that, at the present upper LHC \yy of about 7 TeV,
the $\sigma^{in}/N$ and $\sigma^{tot}/N$ are well above
 the asymptotic values of \eq{asr} and decreasing.  
 The ratios are varying
slowly, for example  at 100 TeV,  $\sigma^{in}/N$ will be about 
0.33, still far from 0.16. It is interesting, however, that 
$\sigma^{el}/N\approx 0.2$
near 7 TeV, is rather closer 
 to the asymptotic
value.

\section{Discussion} \label{disc}

 It is intriguing that the $\sigma^{el}/N$ ratio in
Fig.\,\ref{ratios} is approximately constant and close to the
asymptotic value, even though at these \yys the $\ln^2$ term  in
\eq{nfit} is not yet completely dominant. Apparently, $\sigma^{el}$
and $N$
behave approximately in parallel, 
even well before the aymptoptic region.
 An explanation  for this might be sought  along the following
lines. For the relation
$R\sim \surd N$ to be meaningful, it is necessary to have a
reasonably
well-defined quantity $R$ for the radius of the proton. While this
exists in the black disc limit, at present (LHC) \yys the proton  
has a considerable
``grey'' or semi-transparent area (viewed in impact parameter $b$).

We would like to argue, however, that there is likely a well
defined radius for elastic \sctg  before there is one for the total
\csssn. This is because the elastic and total \sctg at a given $b$ 
are given by the same quantity, $(1-\eta(b))$, squared for elastic
scattering and    
 linear for total cross section.  (In these arguments we take the
\sctg amplitude to 
be purely imaginary, and  $\eta(b)=e^{-\chi(b)}$, with $\chi$ the 
imaginary eikonal.) The quantity $(1-\eta(b))$ is the `opacity' at 
$b$ and varies between an `inner region' where it is close to 1 and
an `edge' where it falls to 0.  Since the square of a number less 
than one is smaller than the number itself, the
transition from $1$ to $0$ will tend to be more abrupt in
the squared, elastic case, and  a radius R will be more well
defined.   This argument  is also supported by the presence of the
diffraction minimum \cite{totem}, which would be washed out if
there were not a relatively well-defined  `edge'. Hence it does not
seem implausible that $\sigma^{el}\sim N$ works for  $\sigma^{el}$ 
before it does for $\sigma^{tot}$.

Assuming that \eq{one} is true, as our results appear to indicate,
it is
naturally possible that it can arise from models other than that of
ref\,\cite{leo}. But in that
 picture for the expanding proton
radius, we can now evaluate the step length $R_0$ in the random
walk from
 our numerical results. This
 would be
$R_0=\sqrt{\frac{\sigma/N}{\pi}}$, using
$\sigma=\pi R^2=\pi R_0^2 N$. From \eq{asr}, using either the
inelastic or
elastic ratio.
\beql{r0}
R_0=\sqrt{\frac{ 0.16\,{\rm  mb}}{\pi}}=0.071 {\rm\  f}\,.
\eeql   
In interpreting this result one should bear in mind that it is
based on using  the \csss per pion. If, as is likely, the pions are
not the primary \pls emitted, but rather come from some fewer
numbers of parent \plsn, then $R_0$ for the primary emission will
be correspondingly larger.
In QCD, where the primary emission would be of gluons, one would
insert a factor for the `number of pions per gluon'.

The uncertainties  due to our estimate
of the \mult via \eq{nfit}
would be reduced if direct measurements of the total \mult
 are obtained from the LHC.
We hope these will become available shortly. Similarly, data from
the full LHC
\yy of 14 TeV will help to test and fix our parameterizations.

\section{Acknowledgments}
 We would like to thank W. Ochs and P. Seyboth for useful
discussions on these matters, and G. Landsberg
  for help with the CMS
data. We would also like to thank the Aspen Center for Physics for
its hospitality, where
this work was supported in part by the
National Science Foundation under Grant No. PHYS-1066293.

\end{document}